# The case for a modern multiwavelength, polarization-sensitive LIDAR in orbit around Mars

Adrian J. Brown[*1], Timothy I. Michaels[1], Shane Byrne[2], Wenbo Sun[3], Timothy N. Titus[4],

Anthony Colaprete[5], Michael J. Wolff[6], Gorden Videen[6]

[1] SETI Institute, 189 Bernardo Ave, Mountain View, CA 94043, USA

[2] Lunar and Planetary Laboratory, 1629 E. University Blvd., Tucson, AZ 85721-0092, USA

[3] Science Systems and Applications, Inc, USA

[4] U.S. Geological Survey, Astrogeology Science Center, Flagstaff, AZ 86001 USA

[5] Space Science Division, NASA Ames Research Center, USA

[6] Space Science Institute, 4750 Walnut St, Suit 205, Boulder, CO, 80301, USA

.

[*] Corresponding author address: Adrian J. Brown, SETI Institute, 189 Bernardo Ave, Mountain View, CA 94043; Tel: +1 (650) 810-0223; email: abrown@seti.org

## HIGHLIGHTS

- ➢ We present the scientific rationale for a modern multi-wavelength, polarization sensitive lidar to be placed in orbit around Mars.
- ➢ Scientific questions that could be addressed focus on the Martian climate and modern-day interactions between surface, ice clouds and dust aerosols.
- ➢ The lessons learned from such an instrument would fundamentally shift our understanding of modern-day volatile transport and deposition. As such, it would have astrobiological implications for past, present and future life on Mars.

**Abstract**

We present the scientific case to build a multiple-wavelength, active, near-infrared (NIR) instrument to measure the reflected intensity and polarization characteristics of backscattered radiation from planetary surfaces and atmospheres. We focus on the ability of such an instrument to enhance, perhaps revolutionize, our understanding of climate, volatiles and astrobiological potential of modern-day Mars.

Such an instrument will address the following three major science themes, which we address in this paper:

**Science Theme 1. Surface**. This would include global, night and day mapping of $H_2O$ and $CO_2$ surface ice properties,
**Science Theme 2. Ice Clouds**. This would including unambiguous discrimination and seasonal mapping of $CO_2$ and $H_2O$ ice clouds, and
**Science Theme 3. Dust Aerosols**. This theme would include multiwavelength polarization measurements to infer dust grain shapes and size distributions.

## 1. Introduction

Our present understanding of the sublimation of surface $H_2O$ and $CO_2$ ices and related atmospheric changes on Mars is the result of recent polewide and seasonal studies of springtime recession using the *CRISM* [1], *Mars Climate Sounder* [2] and *MARCI* [3] instruments on MRO, the *OMEGA* instrument on Mars Express [4,5], the *THEMIS* instrument on Mars Odyssey [6] and the *TES* instrument on Mars Global Surveyor [7]. These investigations have steadily advanced our understanding of major polar processes. However, the confirmed observations of the spatially localized springtime recession phenomena such as geysers (gas/dust jets) [8] and asymmetric retraction of the seasonal cap [9] lead us to ask the key scientific question – what role does spatially localized and temporally intermittent deposition of ices and dust during fall and winter play in the annual $CO_2$ and $H_2O$ cycles which dominate the climate of modern-day Mars?

We discuss herein a proposed instrument called "*Atmospheric/Surface Polarization Experiment at Nighttime*" (ASPEN) *[10]* which is designed in response to this first order scientific question regarding Martian climate.

The *ASPEN* instrument will be a multi-wavelength, altitude-resolved, active near-infrared (NIR) instrument to measure the reflected intensity and polarization characteristics of backscattered radiation from planetary surfaces and atmospheres. The proposed instrument is ideally suited for a mission to Mars to investigate the nature and seasonal abundance of atmospheric dust and icy volatiles, provide insight into surface and cloud/aerosol grain sizes and shapes, evaluate ice and dust particle microphysics and also provide atmospheric column content constituent chemistry during polar night and day.

Previous instruments have given glimpses of cloud and surface ice activity on Mars, but no previous Martian orbital instrument has been able to simultaneously address the following science questions:
a.) Detect clouds up to 100km above the Martian surface during night and day;
b.) Discriminate between $H_2O$ and $CO_2$ ice on the surface and in aerosols in the atmosphere;

c.) Map cloud structure using lidar backscatter and depolarization;

d.) Map large-grained (up to 30cm) $CO_2$ slab ice in the polar night [1,11];

e.) Determine whether the $H_2O$ ice signature in the southern polar trough system is due to cloud [12] or surface ice [13];

f.) Monitor 'cold spot' activity during the polar night and determine whether these enigmatic features are due to $CO_2$ clouds, precipitation or surface ice [14,15];

g.) Monitor night and day gas/dust jet (geyser) activity within the 'Cryptic Region' in southern late winter and early spring and determine what amount of solar energy is required for them to be active [8,16];

h.) Uniquely identify cloud types and platelet/grain orientation, in order to confirm the presence and structure of convective $CO_2$ cloud towers, a potentially critical part of the polar night dynamics and energy partitioning [17];

i.) Provide atmospheric column dust optical depths whenever the instrument is in operation [18,19].

j.) Monitor the spring and summertime retreating polar caps for signs of entrained "sublimation flows" caused by subliming $CO_2$ ice.

k.) Address questions of spatial extent (locality and 'deep transport') of Martian cloud structure, which is anticipated to be on the order of 1km width and is crucial to understanding differences between terrestrial and Martian mesospheric atmospheric dynamics [20,21].

## 2. Instrument Concept and Background

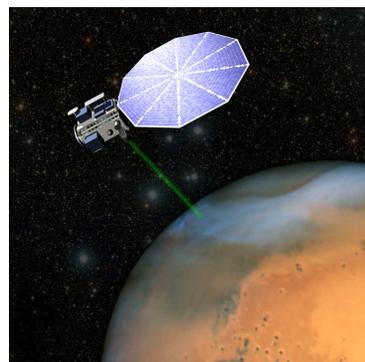

Figure 1. The *ASPEN* instrument in operation at Mars, probing the north polar hood.

The Mars Science community has recognized the need for an *ASPEN*-type instrument. The need for active scanning laser sensors that operate over a range of frequencies was acknowledged in the recent Solar System Exploration Roadmap [22] (page 108). In addition, the Second 2013 Mars Science Orbiter Science Analysis Group (MSO SAG) report stated that a "multibeam lidar" similar to the *LOLA* instrument on Lunar Reconnaissance Orbiter and inheriting many aspects from the *CALIPSO* lidar would "resolve optically dense atmospheric phenomena" and "significantly constrain seasonal mass budgets". In essence, it was thought to be an ideal instrument for a "2013 MSO mission" [23]. A particular emphasis of this MSO SAG report was the need for focussed Polar investigations. In fact, the SAG designated a suite of specific instruments for "P (Polar) type" observations.

A lidar instrument such as *ASPEN* was also recommended in the report on the 3rd International Workshop on Mars Polar Energy Balance and $CO_2$ Cycle [24] and has been emphasized further as a future instrument priority in a white paper submitted to the Planetary Sciences Decadal Survey entitled 'Mars Polar Science for the Next Decade'. Following on after the Mars 2020 rover mission, our targeted mission time frame would be the 2022 launch opportunity and beyond, perhaps as the MICADO Discovery class mission [25].

The scientific impact of such an instrument would be substantial - the Martian climate and its connection to the dynamical environment of the polar nights are unique and poorly understood. Only an active system such as *ASPEN* can adequately investigate the surface and atmospheric characteristics of the Martian polar night.

Figure 1 shows an artistic rendering of the *ASPEN* lidar system deployed in Mars orbit. The eventual spaceflight instrument will be suited for a mission to Mars to investigate the nature and seasonal abundance of icy volatiles, provide insight into surface and cloud grain sizes and geometries, evaluate cloud/aerosol particle microphysics and potentially also provide atmospheric column constituent chemistry during polar night and day.

*Surface spot size and resolution*. Preliminary laser power calculations of common measurement scenarios for the diode pumped fiber laser *ASPEN* instrument estimate the surface spot size at ~25m on the surface and a horizontal resolution of ~275m. This is similar to the resolution achieved by the Nd-YAG *CALIPSO* lidar (Table 1).

| | Laser source | λ (μm) | Laser mJ/pulse | Pulse rate (kHz) | Output pwr (W) | Detector | Mirror diameter/ Field Of View | Mass kg | Power drain W |
|---|---|---|---|---|---|---|---|---|---|
| *Lab ASPEN* | Diode pumped, fiber laser | 1.43-1.67 | 0.04 | 10 | 0.4 | InGaAs APD | 1" optics for source and receiver | ~30 | ~50 |
| *Space ASPEN* | Diode pumped, fiber laser | 1.43-1.67 | 0.04 | 4.5 | 0.18 | InGaAs APD | 80cm/ 0.15mrad | ~15 | ~17 |
| *MOLA* | Diode pumped, Q-switched Nd:YAG | 1.064 | 48 | 0.01 | 0.48 | Si-APD | 50cm/ 0.85mrad | 25.9 | 34.2 |
| *CALIOP* | Diode pumped, Q-switched Nd:YAG | 0.532,1.064 | 110 | 0.02 | 2.2 | Si-APD | 100cm/0.13mrad | 156 | 124 |

Table 1. Comparison of "minimal laboratory" *ASPEN*, eventual spaceborne *ASPEN*, *MOLA* and *CALIOP* capabilities. APD=avalanche photon detector.

*Multiwavelength*. In order to take advantage of the tremendous research and development that has gone into lasers and fiber optic components that operate in the near-IR by the telecommunications industry in recent years, the instrument will operate at wavelengths between 1.43 and 1.67 μm. As we discuss in more detail below, these wavelengths are ideally suited to discriminate $CO_2$ and $H_2O$ ices and vapor using the differential absorption lidar (DIAL) technique originally developed for terrestrial remote sensing [26,27].

These particular characteristics of *ASPEN* are key to generating the type of measurements that will resolve fundamental outstanding questions regarding the Martian climate.

## 2.1. Previous Lidar Mission – MOLA

The highly successful Mars Orbiting Laser Altimeter (*MOLA*) instrument on Mars Global Surveyor measured clouds (see Figure 2) and the height of the seasonal $CO_2$ surface ice accumulations [28–30]. However, its use of a single wavelength (1.064μm) prevents discrimination between $H_2O$ and $CO_2$ clouds using the *MOLA* dataset. In addition, *MOLA* had no ability to determine particle sizes or shapes, nor measure the $H_2O$ or $CO_2$ vapor abundances (Table 1).

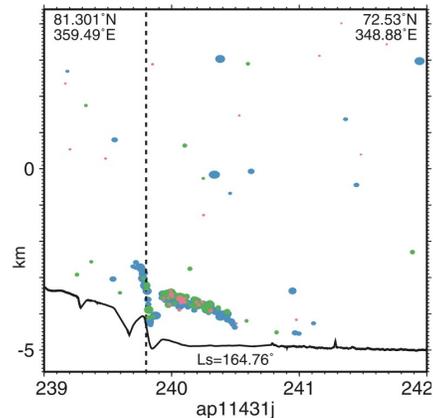

Figure 2. *MOLA* lidar profile over edge of the north polar cap showing "nonground triggers" in colored dots above the solid black line showing the ground elevation. The dotted line is the terminator. $L_s$=164 is northern fall and the north polar hood is responsible for the cloud returns. From [31].

The *MOLA* instrument did demonstrate the ability to detect optically thin Martian dust devils [31]. Consequently, one can have confidence that *ASPEN* will be capable of monitoring dust loading and activity, including that associated with the eruption of 'geysers' in the south polar 'Cryptic' Region [8] - because the *ASPEN* detectors are designed not to saturate over the relatively high albedo Martian ice caps.

*2.2. Previous Lidar Mission – Phoenix*

The *Phoenix* spacecraft landed in the Vastitas Borealis region near the northern pole of Mars (at 68.2 deg N) in May 2008 and operated for 5 months or 152 Martian days (one summer and fall period) [32]. The Phoenix metrology station included a vertical pointing Nd:YAG lidar operating at 1.064 and 0.532 μm. The lidar system successfully detected aerosol structures consistent with Martian cirrus clouds (see Figure 3) and in particular the 'virga' or "Mare's Tails" (ice particles falling from their formation site in the main cloud deck) as they passed over the lander during the local night [33]. Having no polarization capability, the *Phoenix* lidar could not directly determine grain shapes. We consider the *Phoenix* lidar to be a useful pathfinder for more ambitious lidar systems such as *ASPEN*.

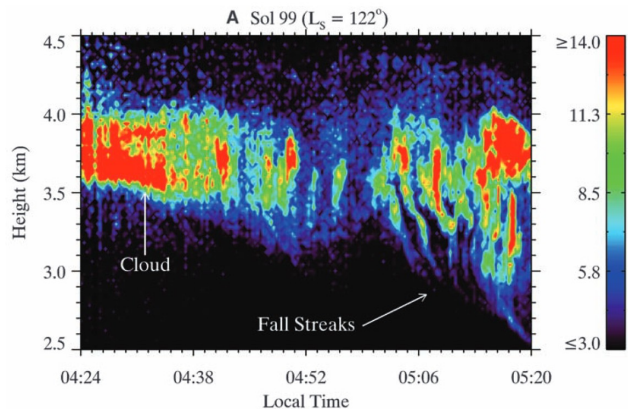

Figure 3. *Phoenix* lidar vertical scan showing fall streaks in northern Martian summer as they pass 4km high over the lander. From [33].

*2.3 Previous Lidar Mission – CALIPSO*

The *CALIOP* laser onboard the Cloud Aerosol Lidar and Infrared Pathfinder Satellite Observations (*CALIPSO*) spacecraft was launched in April 2006 and is still in operation. With an orbit ~700km, it is part of the 'A-Train' of Earth observing satellites. The *CALIOP* laser operates at 1.064 and 0.532 μm, measuring linear polarization in the latter band. The instrument was designed and tested at Ball Aerospace and is operated jointly by NASA and CNES [34].

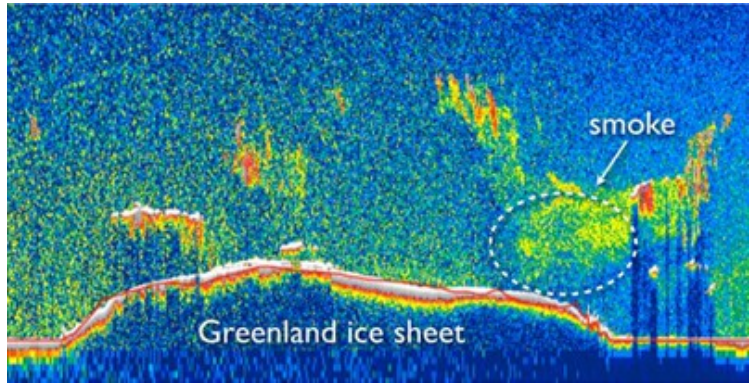

The surface footprint of the *CALIOP* is ~100m and the vertical resolution is 30-60m. Figure 4 shows an example of a *CALIPSO* observation of wildfires over Greenland [35]. The sensitivity to the aerosols associated with the fires provides a clear demonstration of lidar utility for monitoring/characterizing dust and cloud activity across multiple scales, as well as for studies of low lying fogs and sublimation flow events near the Martian surface.

Figure 4. *CALIPSO* satellite vertical scan over Greenland ice sheet showing aerosol soot due to wildfires. From [35].

Although *CALIOP* does not exhibit the same wavelength flexibility and polarimetric capability (i.e. does not measure the full returned Stokes vector) of the *ASPEN* instrument, its enhanced abilities beyond the *MOLA* and *Phoenix* lidar provide further motivation for the *ASPEN* concept of an orbital lidar around Mars.

**3. Scientific Approach to Achieving Planned Objectives**

We now discuss the scientific approach advocated by our team to achieve the Science Objectives of the *ASPEN* instrument. We break the discussion below down by science theme, where each theme embodies at least four key common capabilities:

*Nadir Soundings*. In common with the *MOLA* and *Phoenix* lidars, the *ASPEN* instrument naturally resolves the return pulse from backscattering target materials, including atmospheric constituents, gas and aerosols, clouds and multiple cloud decks and low lying fogs, in addition to the surface return.

*Composition*. The multi-wavelength nature of the instrument allows discrimination of the three major constituents of the Martian volatile cycles – $CO_2$ and $H_2O$ ice and gas, and dust.

*Grain size and distribution*. The Müller matrix polarization capability of the lidar allows determination of the scatterer grain size and place limits on the size distribution.

*Seasonal changes and dynamics*. The orbital nature of the *ASPEN* instrument and mission profile allows us to concentrate on polar observations in order to address key scientific questions that are inaccessible to other instruments. Maps of the changes over the mission lifetime will be key to increasing our understanding and improving our interpretations of the Martian volatile cycles.

To outline the science case, we discuss below a "minimal laboratory" *ASPEN* instrument that could be developed to reduce engineering risks. The "minimal" nature is captured primarily in the cost savings obtained by the use of a small number of laser wavelengths.

*3.1 Science Theme One - Surface*

*3.1.1 Detection of $H_2O$ ice and $CO_2$ ice and discrimination from vapor*

In the 1.4-1.7 μm region there are several $H_2O$ and $CO_2$ ice and gas absorption bands (Figure 5). Table 2 identifies particularly relevant bands for this investigation. Not all of these absorption bands needs to be covered in a "minimal laboratory" instrument, but they may be used by an eventual spaceborne instrument.

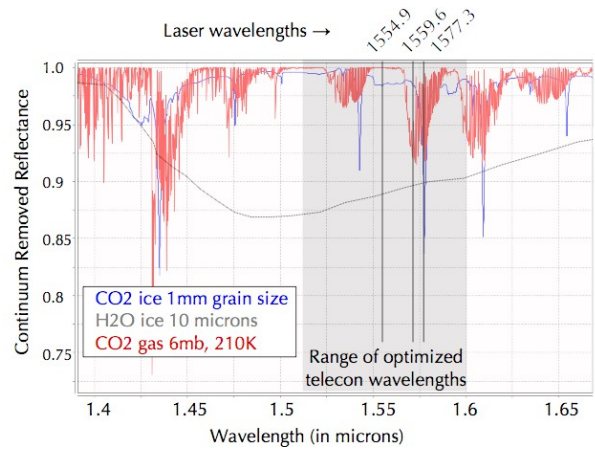

Figure 5. Model reflectance spectra of $CO_2$ vapor (red, normalized to 1) $CO_2$ ice (blue) and $H_2O$ ice (grey). Both ices have been normalized to 1 at 1μm). Black vertical lines show three laser lines to be used for the "minimum" *ASPEN* instrument.

When measured by a coarse spectrometer [36], $CO_2$ ice and vapor lines overlap and therefore can be difficult to separate[37]. However, as seen in Figure 5, when illuminated with a narrow band laser of less than 1 nanometer spectral width, these bands overlap but are separable, particularly because of the nature of narrow $CO_2$ ice bands. To construct Figure 5, we used gas band data from the HITRAN database [38] and $H_2O$ ice optical constants appropriate for 145K [39] and $CO_2$ ice optical constants [40] to model the spectra of $H_2O$ and $CO_2$ ice with grain sizes of 10 and 1000 microns and 40% porosity [9,41], using the albedo model of Shkuratov [42].

*3.1.2 Simulations of the Martian polar surfaces*

Several aspects of the Martian icy polar regions are not well understood. Testing and certifying the *ASPEN* instrument for flight will require preparation of an analog for the Martian $CO_2$ seasonal cap. This will require the marriage of the minimal laboratory instrument with a Martian test chamber. Within the chamber we will construct analog $CO_2$ ice slabs [43] and $H_2O$ and $CO_2$ ice particles to Martian snowpack under a 6-10mbar $CO_2$ atmosphere. This has been done in previous studies, but the amount of dust mixed into the ice surface is not well constrained [44]. During the course of testing this type of instrument, one would anticipate using a range of suspended and surface dust compositions, grain shapes and size distributions to simulate realistic dusty Martian snowpacks and dust-laden atmospheres.

|  | $CH_4$ vapor | $H_2O$ vapor | $H_2O$ ice | $CO_2$ vapor | $CO_2$ ice |
|---|---|---|---|---|---|
| *Wavelength (μm)* | 1.429 | 1.59247 | 1.4-1.75 | 1.44-1.45 | 1.435 |

Table 2. Critical bands for eventual spaceflight ASPEN detection strategy

*3.1.3 Ice identification using multiple wavelengths*

While a "minimal laboratory" instrument might not not include an $H_2O$ vapor band (for example); follow-on work may use a narrow $H_2O$ gas band at 1.59247 μm to obtain column $H_2O$ gas abundance (Table 2). To differentiate ice composition, we will use a ratio of the backscattered reflectance at the absorption band center to that outside the relevant absorption feature. The minimal instrument will use only 3 lasers; therefore, it will only be able to simultaneously detect three phases unambiguously. A ratio of 1.5773/1.5549μm will give $CO_2$ ice, 1.5696/1.5549μm will give $CO_2$ gas and 1.5549/1.5773μm will give the slope of the $H_2O$ ice absorption feature, allowing $H_2O$ ice abundance estimates. More lines would be added in the 1.4-1.7μm region in the eventual orbital instrument to provide simultaneous sensitivity to $H_2O$ and $CO_2$ gas and ice phases (Table 2).

*3.1.4 Surface pressure maps and partial pressure of $H_2O$ vapor*

Using the multiple wavelength DIAL technique discussed above, the *ASPEN* instrument will be able to measure the atmospheric pressure using the $CO_2$ gas band, and partial pressure of $H_2O$ vapor. It will be able to discriminate these from surface ices. Given that $CO_2$ comprises more than 95% of the Martian atmosphere, *ASPEN* can essentially provide the atmospheric surface pressure (generally to 1%). Thus, *ASPEN* will be able to produce global surface total pressure and water vapor partial pressure maps for the entire mission, for each multiple wavelength nadir sounding measurement.

These seasonal maps of surface pressure will be a unique dataset that will be of great value for the Mars climate modeling community, and can be used to generate wind maps, assess atmospheric heat transfer and address questions of current global energy dynamics, including katabatic winds over the polar regions [45], on cap $CO_2$ depositional winds [46] (the "Houben effect"), transfer of $H_2O$ between caps and regolith [47] and countless other Martian atmospheric phenomena.

*3.2 Science Theme Two – Ice Clouds*

*3.2.1 $CO_2$ ice clouds and $CO_2$ ice snowfall*

$CO_2$ ice clouds on Mars were first suggested by Gierasch and Goody [48] and were thought to have been observed by the Mariner 6 and 7 infrared spectrometer [49] although that observation has been disputed [50]. Low brightness temperatures measured by Viking were attributed to $CO_2$ clouds or perhaps snowfall by Forget [51]. Montmessin et al. [52] reported detection of $CO_2$ ice clouds at 100km above the Martian surface using *PFS/SPICAM* in an occultation study.

An analysis of *MCS* observations [2] also indicated substantial cloudiness during the polar night, assumed to be composed of $CO_2$ ice due to the cold atmospheric temperatures. However, being a limb sounder, the *MCS* instrument cannot regularly access the lower 10 -15km of the vertical column of the Martian atmosphere, where the lion's share of atmospheric dust and ice reside. This limitation is the result of the physics of radiative transfer (i.e. multiple scattering effectively obscures any information from this region [53]).

Understanding the distribution of $CO_2$ ice clouds is important because they may have a net surface warming effect [54–56] and may initiate vigorous mixing of the polar night atmosphere, affecting the vertical distribution of temperatures, aerosols and gases [57].

*MOLA* was optimized for topography mapping, not cloud detection. *MOLA* was able to detect clouds that lay within 20km of the surface, and it found clouds mostly on the nightside and in the winter polar hoods of the planet [29,31]. Tantalizingly, *MOLA* found two types of clouds based on height and structure of the return echoes [30]; however, *MOLA* could **not** 1.) detect clouds at greater than 20km altitude; 2.) distinguish definitively between $CO_2$ and $H_2O$ ice; nor 3.) measure albedo or detect presence of gas/dust jets over the seasonal $CO_2$ ice caps (low dynamic range caused saturation).

Nonetheless, coincident *TES* brightness-temperature observations and radio occultation measurements led *MOLA* researchers to suggest the clouds they observed in the polar hoods were most likely $CO_2$ ice clouds [30] and $CO_2$ ice precipitation [58]. MOLA also observed some mid-latitude nighttime clouds that may have been composed of $H_2O$ ice [31]. Confirmation and extension of these observations and identification of $CO_2$ or $H_2O$ clouds is critical to improving our understanding of the Martian thermal budget.

The internal structure of clouds and precipitation streaks were observed over a limited region of the Martian north pole by the *Phoenix* lidar [33] in addition to near-surface fogs [59]. These lidar cloud observations indicate Martian cloud internal structure is likely quite variable and information-rich. Detection of planet-wide four-dimensional (three spatial dimensions and variations with time) cloud decks and mapping of precipitation should be possible using *ASPEN*.

*3.2.2 Cloud nucleation and ice crystal shapes*

There have been numerous attempts to model the grain shapes of $CO_2$ and $H_2O$ ice particles that would be appropriate to Mars, including suggestions that the $CO_2$ ice grains are bipyramidal [60], however observations that would shed light on the true grain shapes have been lacking. As already discussed, *ASPEN* will measure the backscattered Mueller matrix polarization and measure the linear depolarization ratio in order to shed light on the shape (in particular the asphericity) of ice cloud particles as a function of height. The multi-wavelength nature of the observations will also shed light on small (1 micron and smaller) size distributions [61], potentially shedding light on ice nucleation dynamics[62]. Thus, *ASPEN* will be able to address grain sizes and shapes for multiple coincident cloud decks.

*3.3 Science Theme Three - Dust Aerosols*

ASPEN is designed to be sensitive to Martian dust and aerosols. Virga (fall streaks) and precipitation mapping are key parts of our third science theme.

*3.3.1 Polarization Measurements*

The proposed instrument would be the first to obtain polarization measurements of planetary materials under Martian conditions over the 1.4-1.7 μm region. Previous terrestrial studies have proven the utility of polarized light scattered from ice crystals in the atmosphere [63] and on snowpack surfaces [64–66]. *ASPEN* will be capable of measuring the full Stokes vector returned from the linearly/circularly polarized transmitted beam.

*3.3.2 Discriminating dust aerosols and ice particles*

Suspended dust also will have an effect on the detection of ice clouds. This will be particularly true during southern summer dust events. There is evidence that most Martian dust displays a limited size range (average grain sizes of 1.3-1.8 microns [67]) and the polarization returns of these particles will be different from very fine $CO_2$ ice clouds [68] - although some water ice clouds may show similar characteristics. The polar regions experience lower amounts of dust than the rest of the planet [69] however the effect of dust aerosols is of critical importance. The eventual space instrument will discriminate dust from ice using 1.) absence of $CO_2/H_2O$ ice absorption bands, 2.) polarization returns, and 3.) detection height to discriminate airborne dust. Methods have been developed for analyzing the depolarization ratio from *CALIPSO* to characterize the size distribution of airborne dust [70]. *ASPEN's* multi-wavelength capability will enhance the effectiveness of this method.

*3.3.3 Rationale for full Müller Matrix measurements*

The Stokes Vector (Stokes, 1852; van de Hulst, 1957) is used to describe an electromagnetic field $E$, with perpendicular and parallel amplitudes $E_\perp$ and $E_\parallel$ is defined as:

$$\begin{aligned}
I &= \langle E_\parallel E_\parallel^* \rangle + \langle E_\perp E_\perp^* \rangle \\
Q &= \langle E_\parallel E_\parallel^* \rangle - \langle E_\perp E_\perp^* \rangle \\
U &= \langle E_\parallel E_\perp^* \rangle + \langle E_\perp E_\parallel^* \rangle \\
V &= i(\langle E_\parallel E_\parallel^* \rangle + \langle E_\perp E_\perp^* \rangle)
\end{aligned} \qquad (1)$$

where angle brackets indicate a time average and asterisks indicate complex conjugation.

The optical effect of an atmospheric component may then be expressed using its angle-dependent, light scattering Müller matrix as follows:

$$I = M\,I_0 = \begin{pmatrix} M_{11} & M_{12} & M_{13} & M_{14} \\ M_{21} & M_{22} & M_{23} & M_{24} \\ M_{31} & M_{32} & M_{33} & M_{34} \\ M_{41} & M_{42} & M_{43} & M_{44} \end{pmatrix} \begin{pmatrix} I_0 \\ Q_0 \\ U_0 \\ V_0 \end{pmatrix} \qquad (2)$$

where subscript *0* means 'incident' and the Müller matrix of an optical system is represented by elements $M_{11}$ … $M_{44}$. Normal optical remote sensing only measures element $M_{11}$ and passive linear polarization experiments measure $M_{11}$, $M_{21}$ and $M_{31}$. Passive circular polarization experiments measure $M_{41}$. Active Müller matrix measurements, such as the project proposed here, measure all 16 elements of the Müller matrix.

*ASPEN* will measure the backscattered depolarization (degree of linear polarization or DOLP) ratio, which will give further information on particle sizes and shapes. In terms of the returned Stokes matrix, the equation for the DOLP is:

$$DOLP = \frac{\sqrt{(Q^2 + U^2)}}{I} \qquad (3)$$

Spherical or near-spherical hydrometeors (e.g. ice or rain drops) do not depolarize backscattered light; whereas, hexagonal crystals and other shape do. The DPOL ratio for spherical droplets is 0 due to symmetry and typically 0.2-0.8 due to scattering from a variety of asymmetric ice particles [71]. This phenomenon is used in the analysis of *CALIPSO* cloud data to detect oriented ice plates in terrestrial clouds [72]. On Earth, lidar measurements have been used to map the internal structure of clouds where grain shapes change [73–76]. *CALIPSO* data have also been used to map instantaneous connections between cloud vertical structure (via particle orientation) and large-scale climate [77]. This application offers exciting possibilities for the mapping of large scale Martian weather cycles with *ASPEN*.

ASPEN will also create maps of the backscattered degree of circular polarization (DOCP) ratio, which in terms of the returned Stokes matrix is given by:

$$DOCP = \frac{V}{I} \qquad (4)$$

Circular polarization has been shown to decrease with length of travel through a diffuse target (the circular polarization memory effect [78]). Circular polarization measurements are much less common in lidar instruments, but by measuring the circular depolarization of Martian clouds we anticipate to be able to measure their optical depth independently [79] of the degree of returned power, which will put tighter constraints on the inverse problems required to solve for the characteristics of the Martian atmosphere. The degree of circular polarization ratio has also been proposed to help discriminate cloud particle sphericity [80].

Figure 6 displays 16 Müller matrix hemispherical maps for spherical targets produced using the adding-doubling approach [81]. The symmetry of these hemispherical maps allows us to differentiate spherical and non-spherical and Rayleigh scattering target materials[68], and this capability is an important aspect of the *ASPEN* aerosol and cloud mapping approach.

*3.4. Science Matrix*

Table 3 presents the science matrix for the *ASPEN* project. The three scientific themes of the instrument (Surface, Ice Clouds and Dust Aerosols) are linked to project science objectives. The objectives are then addressed individually by the instrument capabilities.

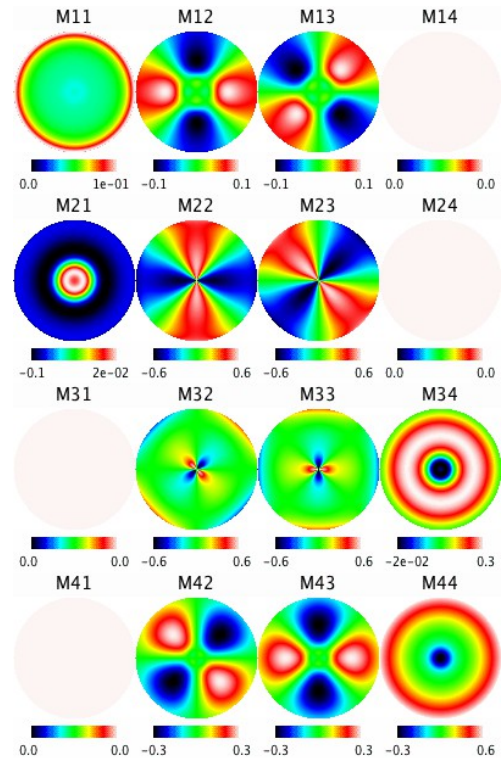

Figure 6. Müller Matrix images of backscattered photons scattered from spherical target modeled by using an adding-doubling radiative transfer model [68]. The laser is incident normal to the target. Color scales indicate intensity relative to the $M_{11}$ element.

| Science Themes | Measurement Objectives | Instrument requirement |
|---|---|---|
| **Science Theme 1. Surface.** To detect, map and quantify deposition of $H_2O$ and $CO_2$ ice during the polar night | 1. Composition. Differentiate surface $CO_2$ ice and $H_2O$ ice | Use NIR laser DIAL technique to differentiate ices |
| | 2. Grain shape and Size. Map $CO_2$ ice and $H_2O$ ice grain size/shape properties | Use DIAL and polarization to map ice properties |
| | 3. Seasonal Changes. Map changes in height as ice is deposited | Use timed laser returns to create high res DTMs to find changes in snow pack height |
| | 4. Seasonal Changes. Determine nature of slab ice south cryptic region [4,8] and re-observe transient "halo" events [87]. | Use NIR laser reflectance DIAL technique to differentiate ices and polarization to determine properties |
| | 5. Nadir soundings. Monitor thermal cold spot activity and determine whether they are due to $CO_2$ snow, $CO_2$ clouds, blizzards or surface ice [51,58,86] | Use NIR laser reflectance DIAL technique to differentiate ices and polarization to determine properties |
| | 6. Surface pressure. Monitor surface pressure and partial pressure of $H_2O$ and produce global, seasonal maps of surface pressure dynamics. | Use NIR laser reflectance DIAL to derive total atmospheric pressure and also $H_2O$ partial pressure. |
| **Science Theme 2. Ice Clouds.** To identify and map fogs, clouds and cloud properties inside and outside the polar hood, on a daily basis. | 1. Nadir soundings. Map cloud heights up to 100km above Martian surface, detect multiple clouds decks | Use NIR reflectance to measure albedo of cloud ice particles |
| | 2. Composition. Determine cloud compositions and find $CO_2$-$H_2O$ ice clusters [85] | Use NIR multiple channel DIAL reflectance technique |
| | 3. Grain shape and Size. Map cloud particle albedo, size and orientation | Use polarization to measure albedo of ice particles |
| | 4. Nadir soundings. Discriminate fogs from $H_2O$ ice deposition on both $CO_2$ ice caps [12,13,46] | Use timed laser returns to discriminate low fogs from surface ice |
| **Science Theme 3. Dust Aerosols.** Map dust storms, precipitation and aerosol loads and particle geometries and orientations on a daily basis. | 1. Nadir soundings. Map ice fall particle heights (virga and precipitation) all mission, determine dust cloud internal structure and multiple decks | Use timed laser returns and strengths to map ice in atmosphere |
| | 2. Grain shape and Size. Map dust aerosols properties | Use NIR polarization to measure particle properties |
| | 3. Seasonal Changes. Map increased dust activity over south pole geysers | Use timed laser returns and full Stokes polarization to detect geyser dynamics and timing |
| | 4. Nadir soundings. Map convective $CO_2$ cloud towers [17] | Use polarization to detect particle orientation / dynamics |

Table 3. Science matrix for the ASPEN project, linking science themes and objectives of the instrument to capabilities of the proposed instrument.

# 4. Implementation Paths

*4.1 Technological Approach and Methodology*

Multi-wavelength laser operations are challenging and will remain so for the foreseeable future. An opportunity exists to use fiber laser amplifier technology developed by the telecommunications industry to make critical measurements in a spaceflight mission.

The currently envisioned spacecraft instrument utilizes multiple diode lasers, each operable at a different wavelength, amplified by a fiber laser stage. The receiver side will consist of a telescope coupled to an indium gallium arsenide (InGaAs) multi-pixel avalanche photo detector (APD). The eventual flight instrument will be scaled to operate at ranges of 250-320km, similar to the MRO orbit. This is around half the altitude of the ~700km altitude *CALIPSO* mission.

*4.2 Martian Mission operations*

As currently envisioned, the *ASPEN* instrument would operate as a line profile instrument, in a similar manner to the *MOLA* lidar [82]. The instrument is best suited for an MGS or MRO-type 250-320km circular orbit but could also operate in an elliptical orbit with reduced sensitivity during apoapsis. For optimized polar measurements, orbital inclination should be between 85 and 92.8 degrees [23]. An elliptical orbit such as that mentioned in the MSO SAG document [83] would allow lidar-occultation measurements of the atmosphere, allowing the atmosphere to be viewed 'side on', thus enabling profile measurements of $CO_2$, $H_2O$ ice and vapor in the Martian atmosphere.

| Planetary Science Decadal Survey Polar Mission Concept Goals *(Calvin et al., 2010)* | Relevant Science Theme of this project |
|---|---|
| 1. Mass, density and volume of seasonal $CO_2$ ice<br>2. Accumulation/ablation rates | 1. Surface |
| 10. Energy exchange during polar night | 2. Clouds |
| 8. Transport of water and dust in and out of polar regions | 3. Aerosols |

Table 4 – Relevance connections with proposed Planetary Science Decadal Survey Missions

Because of its ability to directly detect and discriminate water ice and $CO_2$ ice clouds, the instrument would be directly applicable to four of the ten science goals of the Planetary Science Decadal (PSD) Survey Mission Concept [84] listed in Table 4. The instrument would operate in both day and nighttime conditions, with greater precision during the nighttime due to less reflected background sunlight entering the instrument. During one Martian year of operations, a full summer and winter would be observed at each pole. This would allow an assessment of seasonal cloud and surface volatile activity in addition to monitoring the dynamics of surface and atmospheric pressure and partial pressure of $H_2O$ vapor. This would achieve the top level goals of the instrument and would be considered a complete science mission. An extended mission of an extra year of Martian operations would allow interannual comparisons and additional coverage of the ground surface.

*4.3 Other Planetary Science Missions Applicable for the Proposed Instrument*

We are emphasizing the utility of the *ASPEN* lidar instrument for an orbital Mars mission; however the same type of instrument would be applicable for a range of future missions. It would be ideal for missions to ice covered bodies (e.g. Europa, Enceladus, Triton, even methane ice on Kuiper Belt objects) to investigate the properties of icy surfaces in low sunlight conditions. As part of a Discovery class mission to cometary bodies the system would be ideal for probing the physical properties of a coma. The instrument could also be used in a Venus orbit to probe cloud properties and structure in NIR windows of the Venusian atmosphere.

## 5. Conclusions

We have outlined the science case for a polarization lidar for an eventual orbital mission to Mars. The combination of active, multiple-wavelength measurements with polarimetry makes this instrument concept an essential option in the future inventory of spacecraft instrumentation.

The lessons learned from such an instrument would fundamentally shift our understanding of modern day volatile transport, deposition and would have astrobiological implications for past, present and future life on Mars.

## 6. Acknowledgements

AJB acknowledges support from two grants (NNX11AP23G and NNX13AN21G) from the NASA Planetary Geology and Geophysics program run by Dr. Mike Kelley and two grants from the NASA Mars Data Analysis Program (NNX11AN41G and NNX13AJ73G) administered by Dr. Mitch Schulte.